\documentclass{ptptex}
\usepackage{wrapft}
\usepackage{graphicx}% Include figure files
\usepackage{color}
\usepackage{dcolumn}% Align table columns on decimal point
\usepackage{bm}% bold math
\usepackage{mathrsfs}
\usepackage{dsfont}
\usepackage[colorlinks]{hyperref}
\usepackage{hyperref}
\hypersetup{colorlinks=true,%
            citecolor=blue,%
            filecolor=blue,%
            linkcolor=blue,%
            urlcolor=blue}
\def\t#1{\tilde{#1}}
\def\wh#1{\widehat{#1}}

\def\Sc#1{\textsc{#1}}
\notypesetlogo                       %comment in if to eliminate PTPTeX
\title{%        %You can use \\ for explicit line-break
Reply to ``Comment on `On the Luttinger theorem concerning number of particles in the ground states of systems of interacting fermions','' \href{http://xxx.lanl.gov/abs/cond-mat/0711.3093v1}{arXiv:0711.3093v1}, by A. Rosch%
}

\author{%       %Use \scshape  for the family name
Behnam \textsc{Farid}%
}

\inst{%     %Affiliation, neglected when [addenda] or [errata]
Institute for Theoretical Physics, Department of Physics and Astronomy, University of Utrecht, Leuvenlaan 4, 3584 CE Utrecht, The Netherlands
}

\abst{%         %this abstract is neglected when [addenda] or [errata]
We reply to the Comment by Achim Rosch \citen{AR07} who challenges our finding in Ref.~\citen{BF07} with regard to the validity of the Luttinger theorem in the cases of Mott insulating $N$-particle ground states (even) when the chemical potential used in applying this theorem coincides with the zero-temperature limit of the chemical potential satisfying the equation of state corresponding to $N$ particles. Rosch further argues that the strong-coupling expression for the single-particle Green function presented in Ref.~\citen{AR06} and analyzed in Ref.~\citen{BF07} does not imply destruction of the Mott insulating state at half-filling as a result of an arbitrary weak hopping contribution that breaks particle-hole symmetry and therefore suggests that our conclusion in Ref.~\citen{BF07} to the contrary were incorrect. Here we show the  shortcomings of Rosch's arguments.
}

\begin{document}

\maketitle

\vspace{-10pt}
\noindent {\scriptsize Preprint number: ITP-UU-2007/62}\\

%{\scriptsize{\tableofcontents}}
%%{\footnotesize{\tableofcontents}}
%%\tableofcontents

% I.
%\section{Introduction}
%\label{s1}

In Ref.~\citen{BF07} (Sec.~6.1 herein) we explicitly considered two specific cases which had been investigated by Rosch in Ref.~\citen{AR06}: (1) the local or zero-hopping limit, and (2) the non-local case corresponding to small hopping energy dispersion $\tau_{\bm k}$, $\forall {\bm k}$, in comparison with the effective interaction energy $\t{U}$.
The reader is referred to Ref.~\citen{AR06} for the details of the model to which these cases correspond.

In the local limit, for the Green function\footnote{Throughout we identify $\hbar$ and $k_{\Sc b}$ with unity. As a result of $k_{\Sc b}=1$, in the following $\beta \equiv 1/T$.} $\t{G}({\bm k};z)$ and self-energy $\t{\Sigma}({\bm k};z)$ one has \cite{AR06,BF07}
\begin{equation}
\t{G}_{\rm loc}(z) = \frac{1}{2} \Big( \frac{1}{z - \t{U}/2} +
\frac{1}{z+\t{U}/2} \Big), \label{e1}
\end{equation}
\begin{equation}
\t{\Sigma}_{\rm loc}(z) = \frac{\t{U}}{2} +
\frac{(\t{U}/2)^2}{z}. \label{e2}
\end{equation}
Observing that \cite{AR06,BF07} (below, the last equality applies only for $\vert\mu\vert <\t{U}/2$):
\begin{equation}
\int_{\mu-i\infty}^{\mu+i\infty} \frac{{\rm d}z}{2\pi i}\;
\t{G}_{\rm loc}(z) \frac{\partial}{\partial z} \t{\Sigma}_{\rm
loc}(z) \equiv -\int_{\mu-i\infty}^{\mu+i\infty} \frac{{\rm d}z}{2\pi i}\; \t{\Sigma}_{\rm loc}(z) \frac{\partial}{\partial z} \t{G}_{\rm
loc}(z) = \frac{1}{2}\, \mathrm{sgn}(\mu), \label{e3}
\end{equation}
and given that at zero temperature for $\mu$ inside the Mott gap $(-\t{U}/2,\t{U}/2)$ the number of particles is independent of the value of $\mu$, Rosch \cite{AR06} correctly \cite{BF07} concluded that the Luttinger-Ward identity \cite{LW60}, and thus the Luttinger theorem \cite{LW60},\footnote{In Ref.~\protect\citen{BF07} (Sec. 4.3 herein) we demonstrated that the Luttinger theorem applies \emph{if and only if} the Luttinger-Ward identity is valid, and that deviation from this result implies that the underlying ground state is pathological.} breaks down in the local limit. Although Rosch conceded that the Luttinger theorem applies for $\mu = \mu_{\Sc l}$ (in the local limit, $\mu_{\Sc l}=0$), nonetheless this does not distract from the fact that for insulating ground states no specific aspect in the proof of the Luttinger theorem seems to depend on one distinct value of $\mu$ so that the validity of the Luttinger theorem in the case at hand for $\mu=0$ and its failure for all other $\mu \in (-\t{U}/2,\t{U}/2)$ may be indeed characterised as failure of this theorem \cite{BF07}.

With
\begin{equation}
\zeta_m = \mu + i\omega_m, \label{e4}
\end{equation}
where $\omega_m$, $m=0,\pm 1,\dots$, is the $m$th Matsubara frequency, in Ref.~\citen{BF07} we obtained that
\begin{eqnarray}
\frac{1}{\beta} \sum_{m} \t{G}_{\rm loc}(\zeta_m)\,\frac{\partial \tilde{\Sigma}_{\rm loc}(\zeta_m)}{\partial \zeta_m} &=& \frac{1}{4} \Big[2\tanh\big(\frac{\beta\mu}{2}\big) - \tanh\big(\frac{\beta}{2} (\mu -\frac{\t{U}}{2})\big) -\tanh\big(\frac{\beta}{2} (\mu +\frac{\t{U}}{2})\big)\Big]\nonumber\\
&\equiv& \Phi(\beta\mu,\beta\t{U}). \label{e5}
\end{eqnarray}
The relationship between the sum on the left-hand side (LHS) of Eq.~(\ref{e5}) and the integral on the LHS of Eq.~(\ref{e3}) is established through the equality
\begin{equation}
\lim_{\beta\to\infty} \frac{1}{\beta} \sum_m f_{\beta}(\zeta_m) =
\int_{\mu-i\infty}^{\mu+i\infty} \frac{{\rm d}\zeta}{2\pi i}\;
\lim_{\beta\to\infty} f_{\beta}(\zeta). \label{e6}
\end{equation}
This correspondence, which plays a significant role in the derivation of the Luttinger-Ward identity \cite{LW60}, follows from the consideration that ${\rm d}\zeta_m \equiv \zeta_{m+1} - \zeta_m = 2\pi i/\beta \to 0$ for $\beta\to\infty$, so that in this limit the sum on the LHS of Eq.~(\ref{e6}) is the Riemann sum of the integral on the right-hand side (RHS).

Although the Green function and the self-energy on the LHS of Eq.~(\ref{e5}) are the zero-temperature limits of their corresponding finite-temperature counterparts,\footnote{With reference to Eq.~(\protect\ref{e6}), the LHS of Eq.~(\ref{e5}) corresponds to $\lim_{\beta\to\infty} \frac{1}{\beta} \sum_m f_{\infty}(\zeta_m)$.} we showed in Ref.~\citen{BF07} that as $\beta\to \infty$, for $\mu$ inside $(-\t{U}/2,\t{U}/2)$ and sufficiently far from $\pm\t{U}/2$, to exponential accuracy the RHS of Eq.~(\ref{e5}) coincides with the expression that one would obtain on employing in the LHS of this equation the finite-temperature counterparts of $\t{G}_{\rm loc}(z)$ and $\t{\Sigma}_{\rm loc}(z)$. In fact, for $\beta\to\infty$ to the same accuracy one has \cite{BF07}
\begin{equation}
\frac{1}{\beta} \sum_{m} \t{G}_{\rm loc}(\zeta_m)\,\frac{\partial \tilde{\Sigma}_{\rm loc}(\zeta_m)}{\partial \zeta_m} \sim \Psi(\beta\mu) \equiv \frac{1}{2} \tanh\big(\frac{\beta\mu}{2}\big)\;\;\; \mbox{\rm for}\;\;\; \beta\to\infty. \label{e7}
\end{equation}
The combination $\beta \times \mu$ on the RHS of this expression makes explicit that the zero-temperature limit of the sum on the LHS is approached when $\beta \gg 1/\mu$. Consequently, for $\mu \to 0$ the low-temperature regime corresponds to increasingly larger values of $\beta$. Aside from this fact, one notes that since
\begin{equation}
\lim_{\mu\uparrow 0} \lim_{\beta\to\infty} \Psi(\beta\mu) = -\frac{1}{2},\;\;\; \lim_{\mu\downarrow 0} \lim_{\beta\to\infty} \Psi(\beta\mu) = \frac{1}{2},\;\;\;
\lim_{\beta\to\infty} \lim_{\mu\to 0} \Psi(\beta\mu) = 0, \label{e8}
\end{equation}
the validity or failure of the Luttinger-Ward identity, and thus of the Luttinger theorem, crucially depends on the way in which the zero-temperature limit is effected. In Ref.~\citen{BF07} we referred to the first two \emph{repeated limits} (\S\S~302-306 in Ref.~\citen{EWH27}) in Eq.~(\ref{e8}) as `false', or `spurious', limits.\footnote{Here, as in Ref.~\protect\citen{BF07}, the terms `false' and `spurious' are used in their technical sense and do not connote false or spurious mathematical operations carried out by a particular researcher.} The last repeated limit in Eq.~(\ref{e8}) in conjunction with the expression in Eq.~(\ref{e7}) establish that in the event that the zero-temperature limit of the chemical potential satisfying the equation of state at finite temperatures (i.e. $\mu_{\beta} \equiv \mu(\beta,N,V)$, where $V$ is the volume of the systems in the grand-canonical ensemble under investigation) is vanishing, the Luttinger-Ward identity, and therefore the Luttinger theorem, applies in the local limit on identifying $\mu$ with $\mu_{\infty} \equiv \lim_{\beta\to\infty} \mu_{\beta}$ prior to effecting the zero-temperature limit.

In Ref.~\citen{BF07} (Sec.~6.1.3 herein) we explicitly demonstrated that in the local limit $\mu_{\beta}$ approaches zero faster than $1/\beta$ for $\beta\to\infty$. In consequence of this, not only
\begin{equation}
\lim_{\beta\to\infty} \lim_{\mu\to \mu_{\infty}} \Psi(\beta\mu) \equiv  \lim_{\beta\to\infty} \Psi(\beta\mu_{\infty}) = 0, \label{e9}
\end{equation}
but also
\begin{equation}
\lim_{\beta\to\infty} \Psi(\beta\mu_{\beta}) = 0. \label{e10}
\end{equation}
We have thus explicitly demonstrated \cite{BF07} that, in the local limit the Luttinger-Ward identity, and thus the Luttinger theorem, is obtained on equating $\mu$ with either $\mu_{\infty}$ \emph{or} $\mu_{\beta}$ prior to evaluating the zero-temperature limit. This result unequivocally demonstrates that in the local limit the genuine breakdown of the Luttinger theorem is averted by identifying the value of the thermodynamic variable $\mu$ with $\mu_{\infty}$. In Ref.~\citen{BF07} we present arguments showing that the validity of the Luttinger theorem for $\mu=\mu_{\infty}$ is general and not restricted to the local limit considered here.

As for the non-local cases, to linear order in $\tau_{\bm k}/\t{U}$ for the Green function corresponding to these, Rosch \cite{AR06} obtained that
\begin{equation}
\t{G}({\bm k};z) = \frac{1}{z + \t{U}/2 -\tau_{\bm k} - \t{\Sigma}_{\rm loc}(z)}. \label{e11}
\end{equation}
With
\begin{equation}
\omega_{\pm}({\bm k}) {:=} \frac{\tau_{\bm k}}{2} \pm \frac{1}{2} \big(\tau_{\bm k}^2 + \t{U}^2\big)^{1/2} = \frac{\tau_{\bm k}}{2} \pm \frac{\t{U}}{2} + O\big(\frac{\tau_{\bm k}^2}{\t{U}}\big)\;\;\; \mbox{\rm as}\;\;\; \frac{\vert \tau_{\bm k}\vert}{\t{U}} \to 0 \label{e12}
\end{equation}
denoting the poles of the Green function in Eq.~(\ref{e11}), for the single-particle spectral function $A({\bm k};\omega)$ corresponding to this Green function one has \cite{BF07}
\begin{equation}
A({\bm k};\omega) = \frac{1 - \tau_{\bm k}/(\tau_{\bm k}^2 + \t{U}^2)^{1/2}}{2}\, \delta(\omega - \omega_-({\bm k})) + \frac{1 + \tau_{\bm k}/(\tau_{\bm k}^2 + \t{U}^2)^{1/2}}{2}\, \delta(\omega - \omega_+({\bm k})). \label{e13}
\end{equation}
It is to be noted that the sum of the weights of the two delta functions in this expression is equal to unity, as befits a properly normalised single-particle spectral function.

For a given value of the chemical potential $\mu$,
at zero temperature for the mean value of the number of particles per site per spin spices, $n$, one has
\begin{equation}
n = \frac{2}{\mathcal{N}_{\Sc l}} \sum_{\bm k} \int_{-\infty}^{\mu}
{\rm d}\omega\; A({\bm k};\omega), \label{e14}
\end{equation}
where $\mathcal{N}_{\Sc l}$ is the number of lattice sites and the factor $2$ accounts for the two orbitals per site in the model under consideration. Let now
\begin{equation}
\Omega_- {:=} -\min_{\bm k} \tau_{\bm k} >0,\;\;\;\; \Omega_+ {:=} \max_{\bm k} \tau_{\bm k} >0. \label{e15}
\end{equation}
Introducing the normalised density-of-states function
\begin{equation}
\mathcal{D}(\omega) {:=} \frac{1}{\mathcal{N}_{\Sc l}} \sum_{\bm k} \delta(\omega -\tau_{\bm k}), \label{e16}
\end{equation}
from the expressions in Eqs.~(\ref{e13}) and (\ref{e14}) one trivially obtains that \cite{BF07}
\begin{equation}
n = 1 - \mathcal{C} \;\;\mbox{\rm for}\;\; \max_{\bm k} \omega_-({\bm k}) < \mu < \min_{\bm k} \omega_+({\bm k}) \!\iff\! -\frac{\t{U}}{2}+\frac{\Omega_+}{2} \lesssim \mu \lesssim \frac{\t{U}}{2} - \frac{\Omega_-}{2}, \label{e17}
\end{equation}
where
\begin{equation}
\mathcal{C} {:=} \int {\rm d}\omega\; \frac{\mathcal{D}(\omega)\,\omega}{(\omega^2 + \t{U}^2)^{1/2}}. \label{e18}
\end{equation}
We note that \cite{BF07}
\begin{equation}
\int {\rm d}\omega\; \mathcal{D}(\omega) = 1,\;\;\;\; \int {\rm d}\omega\; \mathcal{D}(\omega)\; \omega = 0. \label{e19}
\end{equation}

For symmetric cases, corresponding to $\mathcal{D}(-\omega) \equiv \mathcal{D}(\omega)$, $\forall\omega$, the constant $\mathcal{C}$ is identically vanishing. For asymmetric cases and $\t{U} > \max(\Omega_-,\Omega_+)$ one has the following uniformly convergent series \cite{BF07}
\begin{equation}
\mathcal{C} = \frac{1}{\t{U}}\, \mathcal{I}_1 - \frac{1}{2\t{U}^3}\, \mathcal{I}_3 + \frac{3}{8\t{U}^5}\, \mathcal{I}_5 -\dots,\;\;\;
\mbox{\rm where}\;\;\; \mathcal{I}_j {:=} \int {\rm d}\omega\; \mathcal{D}(\omega)\, \omega^j. \label{e20}
\end{equation}
On account of the second expression in Eq.~(\ref{e19}) one has $\mathcal{I}_1 =0$, so that $\mathcal{C}$ is at the largest of the order of $(\omega_0/\t{U})^3$ where $\omega_0 \in (-\Omega_-,\Omega_+)$; for convenience, below we shall refer to $\mathcal{C}$ as being at the largest of the order of $(\tau_{\bm k}/\t{U})^3$. \emph{Note that a $\mathcal{C} \not=0$ implies a net amount of total spectral weight transfer from the band $\omega_{\mp}({\bm k})$ to the band $\omega_{\pm}({\bm k})$ when $\mathcal{C} \gtrless 0$.}

Since for asymmetric cases $\mathcal{C}$ is not identically vanishing (it may however be vanishing for some \cite{BF07} discrete values of $\t{U}$), from Eq.~(\ref{e17}) it follows that for $\mu$ in the interval given in Eq.~(\ref{e17}) the ground state of the system under consideration \emph{cannot} be half-filled, i.e. $n=1$ is ruled out; the requirement for the ground state to be half-filled, turns the ground state metallic \cite{BF07}.

Rosch \cite{AR07} considers the conclusions in the previous paragraph as incorrect. He argues that the Green function presented in Eq.~(\ref{e11}) being correct only to linear order in $\tau_{\bm k}/\t{U}$, $\mathcal{C}$, which at the largest is of the order of $(\tau_{\bm k}/\t{U})^3$ (see above), should be identified with zero. We, on the other hand, reason that the half-filled metallic ground states predicted by the Green function in Eq.~(\ref{e11}) being induced by merely $\mathcal{C}\not=0$, irrespective of how small $\vert\mathcal{C}\vert$ may be, signifies the fact that these half-filled metallic ground states are not to be taken as physically viable. In the closing part of Sec.~6.1.5 in Ref.~\citen{BF07} we stated that:
\begin{itemize}
\item[{}] {\small ``Summarising, since not until the third order in $\tau_{\bm k}/\t{U}$ can $\mathcal{C}$ deviate from zero in asymmetric cases, we have thus the clearest evidence that \emph{first-order results are in principle incapable of establishing break-down of the Luttinger theorem in asymmetric cases}.''}
\end{itemize}
In Ref.~\citen{BF07} (the closing paragraph of Sec.~6.1.13 herein) we further acknowledged the observation by Rosch \cite{AR06} that for $\mu$ in the vicinity of $\pm\t{U}/2$ the leading-order perturbation expansion resulting in the Green function in Eq.~(\ref{e11}) becomes inadequate.

Now let us investigate the consequences of Rosch's proposal of identifying $\mathcal{C}$ with zero. Since (cf. Eq.~(\ref{e20}))
\begin{equation}
\frac{\tau_{\bm k}}{(\tau_{\bm k}^2 + \t{U}^2)^{1/2}} = \frac{\tau_{\bm k}}{\t{U}} -\frac{1}{2} \big(\frac{\tau_{\bm k}}{\t{U}}\big)^3 + \frac{3}{8} \big(\frac{\tau_{\bm k}}{\t{U}}\big)^5 - \dots\;\;\; \mbox{\rm for}\;\;\; \frac{\vert\tau_{\bm k}\vert}{\t{U}} < 1, \label{e21}
\end{equation}
it follows that identification of $\mathcal{C}$ with zero amounts to approximating $\tau_{\bm k}/(\tau_{\bm k}^2 + \t{U}^2)^{1/2}$ by the leading-order term $\tau_{\bm k}/\t{U}$. Following Eq.~(\ref{e13}), this approximation combined with the concomitant leading-order approximations of $\omega_{\pm}({\bm k})$ result in the approximate single-particle spectral function
\begin{equation}
A'({\bm k};\omega) \equiv \frac{1 -\tau_{\bm k}/\t{U}}{2}\, \delta(\omega - \tau_{\bm k}/2 + \t{U}/2) + \frac{1 +\tau_{\bm k}/\t{U}}{2}\, \delta(\omega - \tau_{\bm k}/2 - \t{U}/2). \label{e22}
\end{equation}
One readily verifies that this function corresponds to the approximate single-particle Green function
\begin{equation}
\t{G}'({\bm k};z) = \frac{1}{2} \Big(\frac{1}{z - \tau_{\bm k}/2 +\t{U}/2} + \frac{1}{z - \tau_{\bm k}/2 -\t{U}/2}\Big) \equiv \frac{1}{z -\tau_{\bm k} +(\tau_{\bm k}^2 -\t{U}^2)/(4 z)}. \label{e23}
\end{equation}
With $\t{G}_0({\bm k};z) = 1/(z-\tau_{\bm k} + \t{U}/2)$, from the Dyson equation one immediately infers that $\t{G}'({\bm k};z)$ corresponds to the self-energy
\begin{equation}
\t{\Sigma}'({\bm k};z) \equiv  \t{\Sigma}_{\rm loc}(z) -\frac{\tau_{\bm k}^2}{4 z}. \label{e24}
\end{equation}
In contrast to $\t{G}({\bm k};z)$, $\t{G}'({\bm k};z)$ is thus seen to correspond to a \emph{non-local} self-energy. It may be instructive to express $\tau_{\bm k}^2/z$ as $(\tau_{\bm k}/z)\, \tau_{\bm k}$ whereby one may view $\tau_{\bm k}^2/z$ as `linear' in $\tau_{\bm k}$ with a diverging coefficient for $z\to 0$. This observation is noteworthy in that the leading-order hopping contribution to $\t{G}_{\rm loc}(z)$, resulting in the Green function in Eq.~(\ref{e11}), is the $-\tau_{\bm k}$ encountered in the denominator of the expression on the RHS of this equation.

Not surprisingly, one can explicitly show that the half-filled ground state corresponding to $\t{G}'({\bm k};z)$ is Mott insulating, exactly as is the case for $\t{G}_{\rm loc}(z)$ and $\t{\Sigma}_{\rm loc}(z)$, and indeed for $\t{G}{\bm k};z)$ and $\t{\Sigma}_{\rm loc}(z)$ when $\mathcal{D}(\omega)$ is symmetric \cite{BF07}. This interesting observation notwithstanding, it should be evident that the considerations in Ref.~\citen{BF07} have \emph{no} bearing on $\t{G}'({\bm k};z)$, but on the $\t{G}({\bm k};z)$ presented in Eq.~(\ref{e11}).

We have thus unequivocally established that a theoretical framework in which $\mathcal{C}$ is identified with zero, on account of it being of higher order than linear in the small parameter $\tau_{\bm k}/\t{U}$, is strictly distinct from the framework in which the $\t{G}({\bm k};z)$ in Eq.~(\ref{e11}) is the exact Green function. In this light, we believe that Rosch \cite{AR07} misinterprets a number of matters by stating that \cite{AR07}
\begin{itemize}
\item[{}] {\small ``His [Farid's] argument is based on a surprising result of his calculations: he claims [3] that an arbitrary small breaking of particle-hole symmetry transforms the half-filled Mott insulator into a metal, or, equivalently, that the particle-hole asymmetric system in not half-filled if the chemical potential is located within the gap! In our opinion, this is obviously wrong. For example, it contradicts the observation that small perturbations have no effects in systems with a finite gap (in the two-band Mott insulator under consideration both the charge and the spin gap are finite). In the appendix we sketch the formal argument which can be used to prove this.''}
\end{itemize}
The ``surprising'' result to which Rosch \cite{AR07} refers is presented in Eq.~(\ref{e17}) above, which arises from a rigorous treatment of the Green function in Eq.~(\ref{e11}). To contrast this result, corresponding to an \emph{approximate} Green function, with one deduced from \emph{exact} considerations, and subsequently characterise the calculations leading to the ``surprising'' result as incorrect is not only logically unsound, but is grossly unfair. Nowhere in Ref.~\citen{BF07} have we indicated or implied that the conclusions that Rosch identifies as ``surprising'' and ``not correct'' should be viewed as exact in the absolute sense; the validity of these conclusions is relative to a framework in which the Green function in Eq.~(\ref{e11}) is held as exact.

Similarly, the statement with regard to the stability of the half-filled Mott insulating phase against small asymmetric perturbations cannot have any place within the framework where the Green function in Eq.~(\ref{e11}) is exact. The exact considerations presented in the appendix to Ref.~\citen{AR07}, which are similar to those explicating the robustness of the Mott phase against small-amplitude hopping terms in the Bose-Hubbard Hamiltonian (\S~10.1 in Ref.~\citen{SS99}), does not apply here; such considerations \emph{in principle} apply only if the Green function under consideration is the exact Green function. Stated differently, with $\wh{H}$ denoting the Hamiltonian of the system under investigation, it is required that $\t{G}({\bm k};z)$ correspond to the exact $N$-particle ground state of $\wh{H}$. As it stands, the $\t{G}({\bm k};z)$ in Eq.~(\ref{e11}) not only does not correspond to the $N$-particle ground state of $\wh{H}$, with absolute certainty it does not correspond to the $N$-particle ground state of \emph{any} interacting Hamiltonian operating in infinite-dimensional $N$- and $N\pm 1$-particle Hilbert spaces.

To summarise, those conclusions arrived at in Ref.~\citen{BF07} with which Rosch \cite{AR07} specifically disagrees, \emph{are} exact. The source of Rosch's contention appears to be two-fold. Firstly, Rosch believes, unwarrantedly as our above calculations show, that the constant $\mathcal{C}$ should be identified with zero. Secondly, for reasons that are not apparent to us, Rosch \cite{AR07} asserts that the ``surprising'' result in Eq.~(\ref{e17}) is not to be trusted on account of it contradicting the robustness of the Mott insulating state against small hopping terms in the underlying Hamiltonian. As for the first point raised by Rosch, we have explicitly shown that identification of $\mathcal{C}$ with zero leads to transformation of the $\t{G}({\bm k};z)$ in Eq.~(\ref{e11}) into an entirely different Green function, i.e. the $\t{G}'({\bm k};z)$ in Eq.~(\ref{e23}); evidently, the considerations in Ref.~\citen{BF07} have no bearing, whatever, on the properties of $\t{G}'({\bm k};z)$. As for Rosch's second point, we have emphasised that our ``surprising'' finding is only so by assuming that the Green function in Eq.~(\ref{e11}) were exact in the absolute sense. We have further underlined the fact that the considerations in the appendix to Ref.~\citen{AR07} would apply if the Green function in Eq.~(\ref{e11}) corresponded to the exact ground state of the interacting Hamiltonian under considerations.

\vspace{8pt}

I should like to thank A. Rosch for kindly providing me with a draft of his Comment in advance of its publication.

\end{document}